\documentclass[%
 aip,
 amsmath,amssymb,
 reprint,%
]{revtex4-1}

\usepackage{graphicx}

\usepackage{color}
\usepackage{url,hyperref,lineno}
\usepackage{dcolumn}
\usepackage{bm}

\usepackage[utf8]{inputenc}
\usepackage[T1]{fontenc}
\usepackage{mathptmx}

\begin{document}

\preprint{AIP/123-QED}

\title[Experimental study of speckle patterns]{Experimental study of speckle patterns generated by low-coherence semiconductor laser light}

\author{D. Halpaap}
\affiliation{ 
Departament de Fisica, Universitat Politecnica de Catalunya, St. Nebridi 22, 08222 Terrassa, Barcelona, Spain
}%

\author{M. Marconi}
\affiliation{ 
Centre de Nanosciences et de Nanotechnologies, CNRS, Universite Paris-Sud, Universite Paris-Saclay, 91405 Orsay cedex, France
}
\altaffiliation[Now at ]{Institut Physics de Nice (INPHYNI) and CNRS UMR, Universite Cote D'Azur, France}

\author{R. Hernandez}%
\affiliation{ 
Centre de Nanosciences et de Nanotechnologies, CNRS, Universite Paris-Sud, Universite Paris-Saclay, 91405 Orsay cedex, France
}%

\author{A. M. Yacomotti}%
\affiliation{ 
Centre de Nanosciences et de Nanotechnologies, CNRS, Universite Paris-Sud, Universite Paris-Saclay, 91405 Orsay cedex, France
}%

\author{J. Tiana-Alsina}
\author{C. Masoller}
\email{cristina.masoller@upc.edu}
\affiliation{%
Departament de Fisica, Universitat Politecnica de Catalunya, St. Nebridi 22, 08222 Terrassa, Barcelona, Spain
}%

\date{\today}

\begin{abstract}
Speckle is a wave interference phenomenon that has been studied in various fields, including optics, hydrodynamics and acoustics. Speckle patterns contain spectral information of the interfering waves, and of the scattering medium that generates the pattern. Here we study experimentally the speckle patterns generated by the light emitted by two types of semiconductor lasers: conventional laser diodes, where we induce low-coherence emission by optical feedback or by pump current modulation, and coupled nanolasers. In both cases we analyze the intensity statistics of the respective speckle patterns to inspect the degree of coherence of the light. We show that that speckle analysis provides a non-spectral way to assess the coherence of semiconductor laser light.
\end{abstract}

\maketitle

\begin{quotation}
Semiconductor laser diodes are widely used as illumination light sources in imaging applications. These lasers are popular because they are low-cost, emit a stable output and cover a wide range of wavelengths. However, illumination with coherent laser light produces a spatial interference pattern, known as speckle pattern, which degrades the image quality. On the other hand, speckle patterns contain useful information about the spectral properties of the light.  Speckle patterns can also be used to infer properties of the scattering medium that generates the speckle.  Here we study experimentally the speckle patterns generated by two different types of semiconductor lasers: a conventional laser diode, in which we induce low-coherence emission by periodic temporal perturbations, and a nanolaser, whose low-coherence emission is mainly due to noise and coupling to another nanolaser. In both cases we show how the speckle intensity statistics changes due to the different degrees of coherence of the light.
\end{quotation}

\section{Introduction}

Optical speckle is a wave interference phenomenon that is produced by coherent light, i.e., a light beam emitted by a laser that has a well-defined wavefront phase relation~\cite{libro}. Laser light can be temporal and/or spatially coherent. Temporal coherence, used for example in a Michelson interferometer, refers to the possibility that a laser beam coherently interferes with a time-delayed version of it self. The degree of temporal coherence is measured by the light's spectral width. On the other hand, spatial coherence refers to the possibility that two spatially shifted laser beams interfere. The classical Young's double slit experiment is based on spatial coherence, and also, on temporal coherence.

\begin{figure}[!t]
\includegraphics[width=1.0\columnwidth]{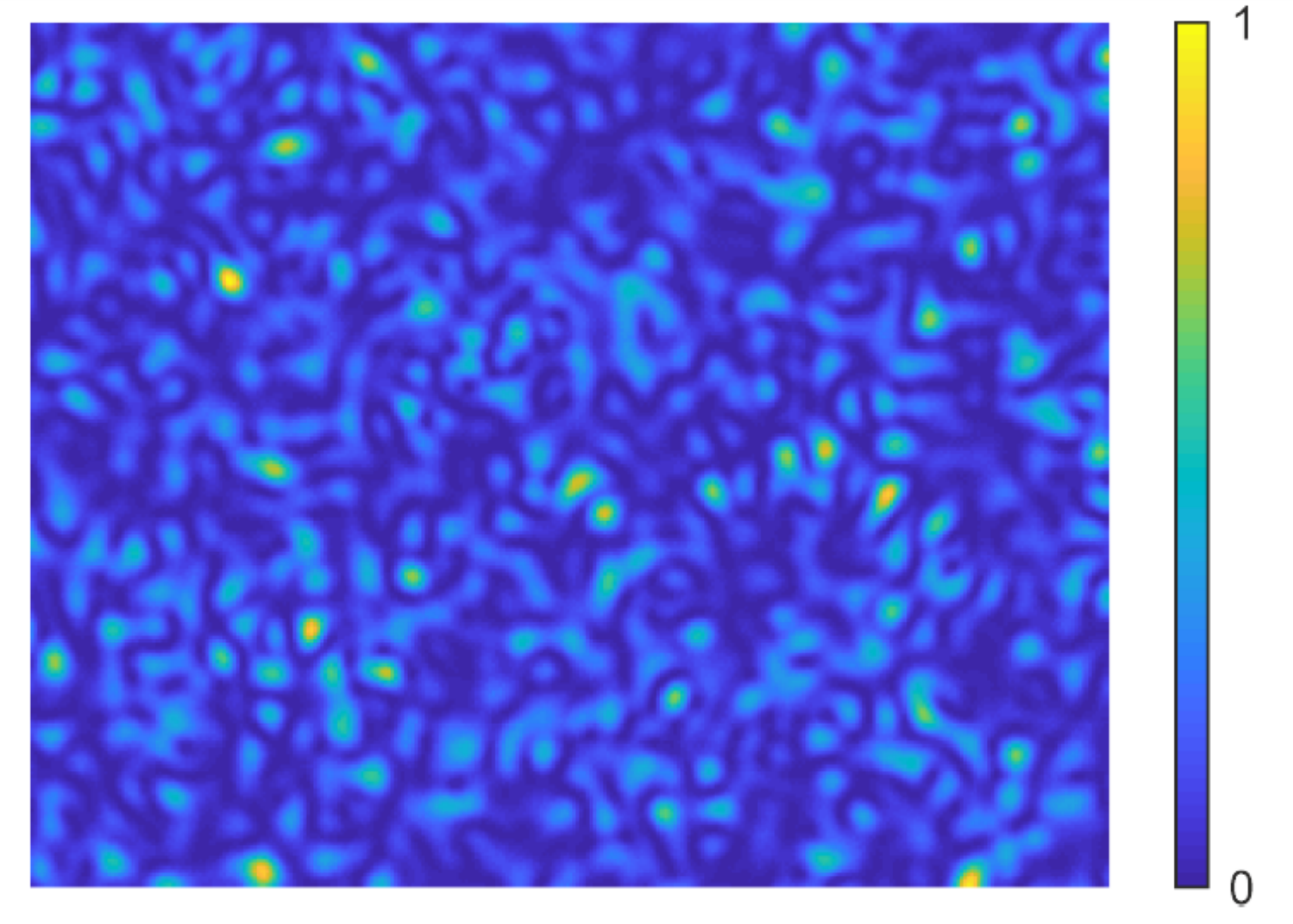}
\caption{Example a speckle pattern that originates from laser diode light that is transmitted through a ground glass and is directly imaged by a digital CCD camera. The characteristic area of the individual speckle spots is larger than the area of several pixels.
The color code indicates the intensity levels of the black and white frame (colored for better visibility). \label{fig1}}
\end{figure}

The granular, noise-like structure of a speckle pattern (such as that shown in Fig.~\ref{fig1}) is a problem when using laser light for imaging. For example, speckle degrades the image quality of laser-based cinema projector systems~\cite{guy}, and efforts have focused on strategies to reduce or mitigate speckle~\cite{guy2}. New laser cavity geometries and alternative feedback mechanisms have been proposed to reduce the spatial coherence of laser light~\cite{hui_review}. Random lasers (lasers made from disordered materials that trap light via multiple scattering) specifically engineered to have low spatial coherence have been shown to provide speckle-free full-field imaging~\cite{hui_nat_phot}. In Ref.\cite{hui_nat_phot} the active medium was a dye solution interspersed with scattering particles, whose emission characteristics  could be tuned by adjusting the scattering strength and the pump geometry. However, such lasers are not convenient light sources for imaging applications such as laser projectors or microscopes because they are neither compact, nor electrically pumped. A compact, electrically pumped light source able to provide both, low spatial coherence  (needed for speckle suppression) and high brightness (needed for high-speed imaging) was demonstrated in Ref.~\cite{hui_pnas}, where a semiconductor laser was fabricated with a chaotic, D-shaped cavity.  In a chaotic cavity (such as a two-dimensional billiard) the ray dynamics is chaotic. Such cavity can allow a huge number of spatially independent modes to lase simultaneously. In Ref.~\cite{hui_pnas} the D-shaped cavity was designed to maximize the number of lasing modes (over 1000 modes could lase simultaneously). Extreme multimode emission provided the desired high brightness and low coherence required for fast speckle-free imaging.

On the other hand, because speckle patterns result from the interferometric summation of multiply scattered wavefronts, they contain useful information of the scattering medium that produces the speckle, and can therefore be used for imaging and sensing~\cite{review_imaging,avalanche_pre,pnas_2017,optica_2018}. For example, speckle has been used for remote vibration analysis and the remote recovery of audio signals~\cite{sensing}. Different approaches for extracting meaningful information from speckle patterns have been proposed~\cite{rsi_2005,method,oe2020}.

In addition, the intensity distribution, the speckle spot size, and speckle correlations provide information of the spectral properties of the source that generates the light~\cite{chaotic_laser_pre}.  High-precision spectrometers have been demonstrated, where wavelength-dependent speckle patterns are detected and, after calibration, used to recover the spectrum of the light~\cite{cao,dholakia}.

Here we study experimentally the speckle patterns generated by two different types of semiconductor lasers, whose emission has a low degree of temporal coherence that originates from different physical phenomena. We consider a conventional diode laser, with temporal dynamics induced by optical feedback or by current modulation (well-known sources of optical instabilities and chaos~\cite{book_ohtsubo,carlos}), and we also investigate a novel type of nanolaser, whose low-coherence emission is  due to noise and evanescent coupling to another nanolaser~\cite{nano1,nano2}. 

We use a multimode fiber (for the laser diode) and a ground glass disk (for the nanolaser) to generate speckle, and two digital cameras (for the laser diode, a camera sensitive in the visible range; for the nanolaser, a camera sensitive in the near-IR) to record images of the speckle patterns. Then, we analyze the probability distribution function of the pixel intensity values, and measure the amount of speckle in the image with the {\it{speckle contrast}}~\cite{libro},
$C={\sigma_I}/{\left< I \right>}$,
where $\left< I \right>$ and $\sigma_I$ are the mean value and the standard deviation of the intensity distribution. 

The intensity distribution can be described by considering speckle as the result of a ``random walk''~\cite{libro}, i.e. the summation of many phasors (having both an amplitude and a phase) with random amplitudes and phases in the complex plane. Depending on the phases, the components add up (or interfere) constructively or destructively to a total amplitude. Assuming that the phases are uniformly distributed in [0,$2\pi$) and that the amplitudes are uncorrelated~\cite{cao_prl}, the distribution of field amplitudes is the Rayleigh distribution, which results in a negative exponential intensity distribution (so called thermal light). Speckle patterns with this intensity distribution are called fully developed and have $C=1$. On the other hand, when $C\sim0$ there are no speckles, i.e., there is no coherent wave interference because the light is fully incoherent. If  $C>1$, the intensity distribution is non-thermal. 

Studying the turn-on of a free-running laser diode (without optical feedback or current modulation) some of us have recently shown\cite{oe2019} that, as the laser pump current increases, when coherence emerges (i.e., at the lasing threshold) $C$ grows abruptly, from very low values below the threshold (where the emission is dominated by spontaneous emission) to high values above the threshold (where the emission is dominated by stimulated emission) and reaches a plateau at a high $C$ value (if the emission is fully coherent, $C = 1$).

Here, we consider different types of lasers (conventional laser diodes and nanolasers) that emit light that is partially incoherent, and show that speckle analysis consistently provides a non-spectral way to assess the level of coherence of the laser light.

This paper is organized as follows: Sec. 2 presents the experimental setup used to study the speckle generated by a laser laser with temporal dynamics induced by either optical feedback or by current modulation;  Sec. 3 presents the results obtained with this setup; Sec. 4 presents the setup used to study speckle generated by coupled nanolasers and Sec. 5  presents the results obtained; Sec. 6 presents a discussion and our conclusions. 

\section{Laser diode setup}

\begin{figure}[!t]
\includegraphics[width=1.0\columnwidth]{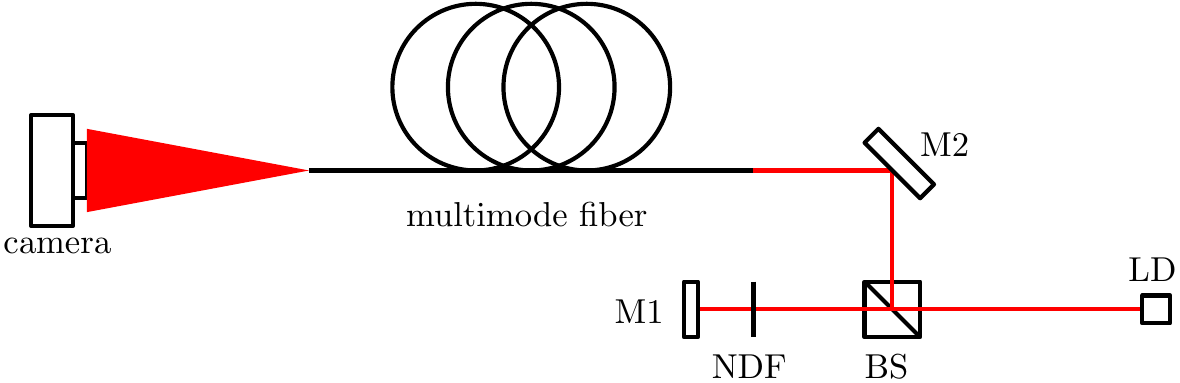}
\caption{Experimental setup to study speckle generated by the light of a diode laser with temporal dynamics induced by optical feedback or pump current modulation. The optical density of the neutral density filter (NDF) and the transmission of the beam splitter (BS) determine the amount of feedback. M1 and M2 are mirrors that provide optical feedback. \label{fig2}}
\end{figure}

Figure~\ref{fig2} displays the experimental setup. We use a laser diode (Hitachi HL6501MG) emitting at 656~nm, with threshold current 39.5~mA. A mirror (M1) in the laser beam path defines a 50~cm long external cavity used to introduce optical feedback. A beam splitter (BS) is used to couple light out for measurements.  The optical density of the neutral density filter (NDF) and the transmission of the BS determine the feedback strength, which was quantified by the feedback-induced threshold reduction to be 16.4\%. The laser beam is sent through a step-index multimode optical fiber with a core diameter of 200 $\mu$m (Thorlabs M72L02) and the light exiting at the other end is imaged by a 8-bit CMOS camera (IDS UI-1240SE-M). Speckle is created by the interference of different guided modes in the fiber. We also measure the spectrum of the light using an optical spectrum analyzer (OSA, Anritsu MS9710C).

To test the robustness of the results, we perform a second set of experiments in which there is no optical feedback, but the pump current of a diode laser (Thorlabs HL6750MG, emitting at 685~nm with threshold current 26.7~mA) is modulated with a square wave signal (provided by an arbitrary waveform generator, Agilent 81150A), such that the laser is turned off and on in each cycle (the dc value of the pump current is 30~mA and signal amplitude is $\pm15$~mA). As in the feedback experiment, speckle is created inside the multimode optical fiber and imaged with the camera. At the same time, the spectrum is recorded by the OSA. The pump current modulation frequency is 200~Hz and the exposure time of the CCD camera is 50~ms. In this way, each frame captures the image of the speckle pattern, averaged over four modulation cycles. To test the influence of the modulation frequency, we also performed measurements with a modulation frequency of 200~kHz. 

\section{Laser diodes results}

\begin{figure}[!t]
\includegraphics[width=1.0\columnwidth]{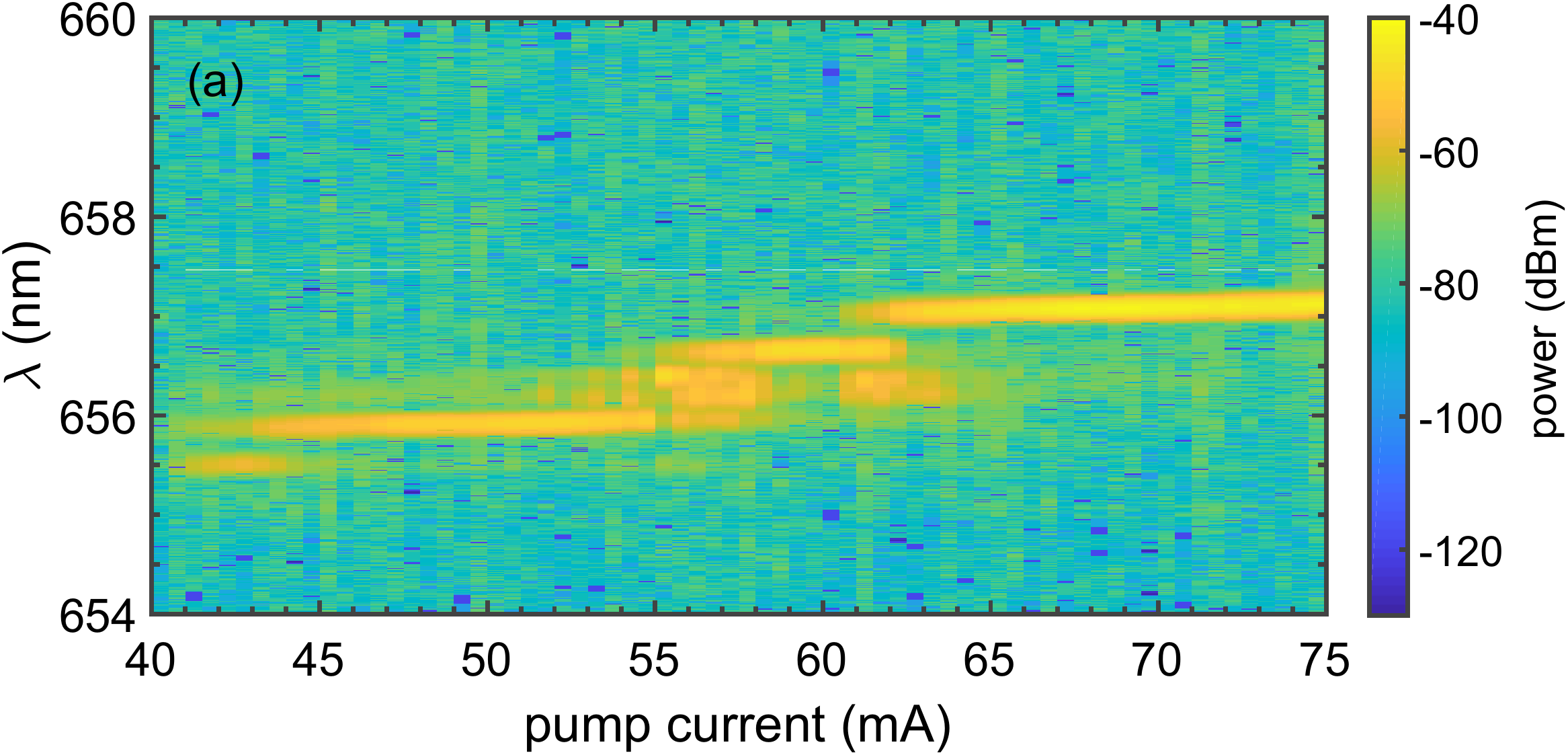}
\includegraphics[width=1.0\columnwidth]{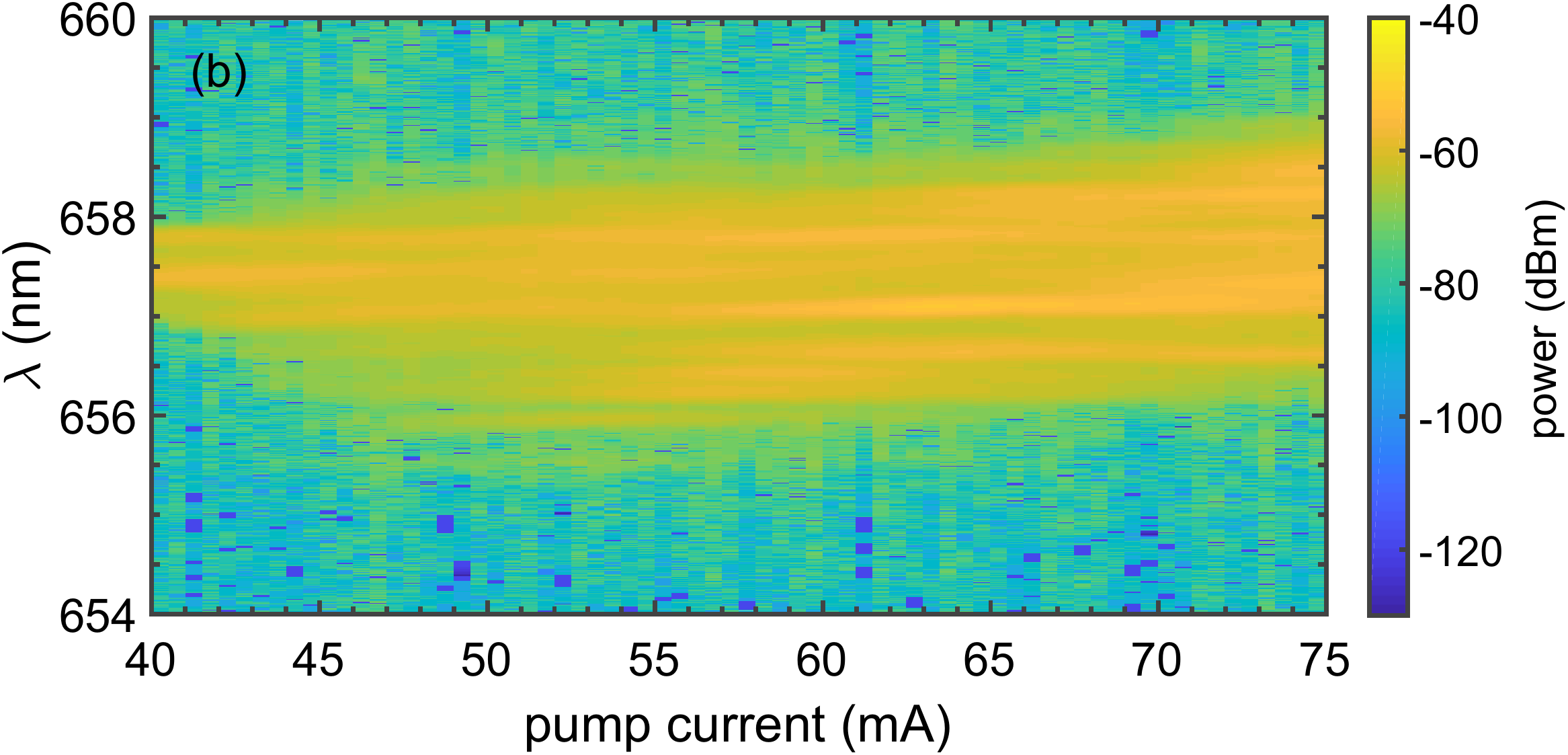}
\includegraphics[width=1.0\columnwidth]{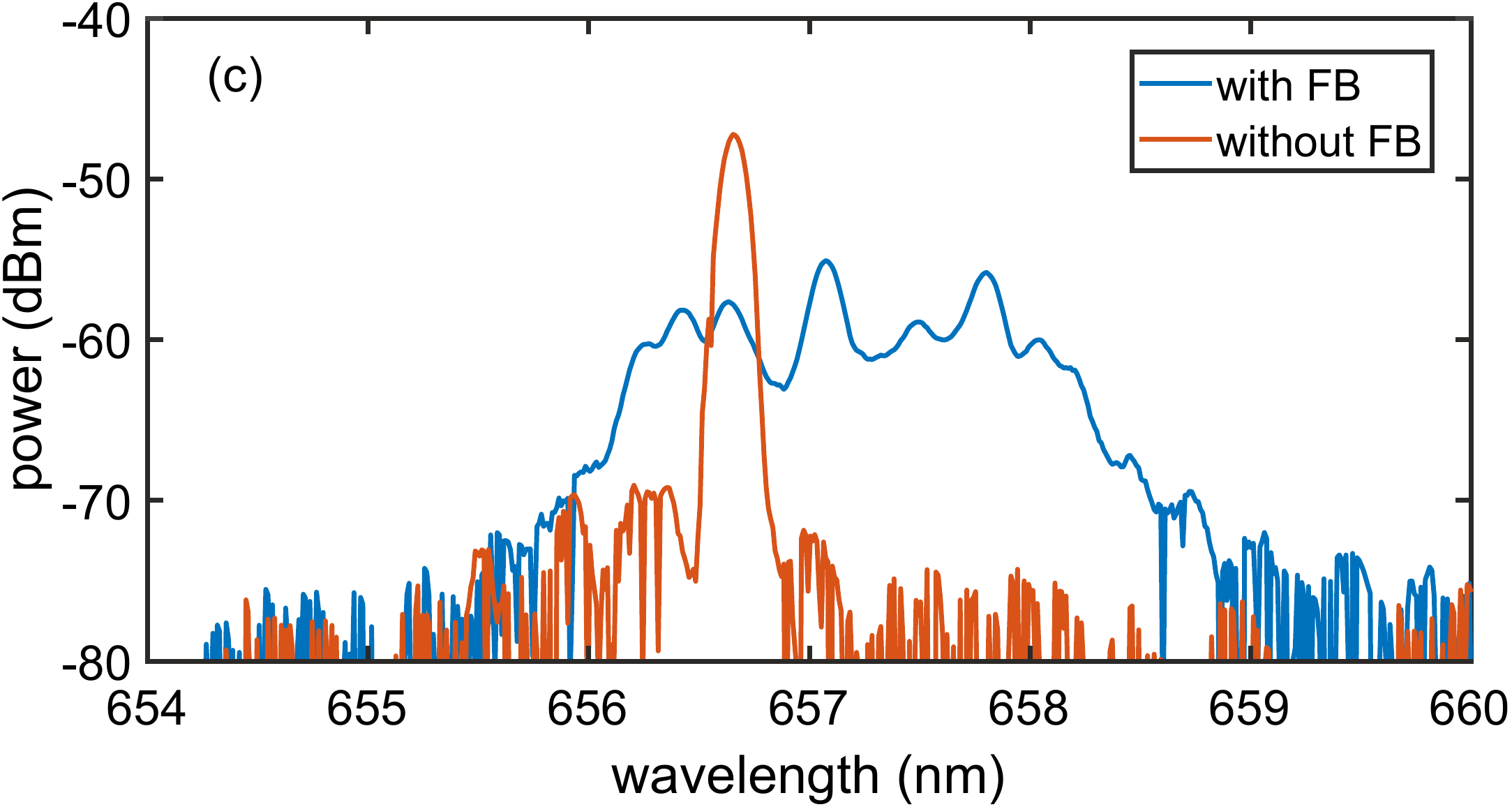}
\includegraphics[width=1.0\columnwidth]{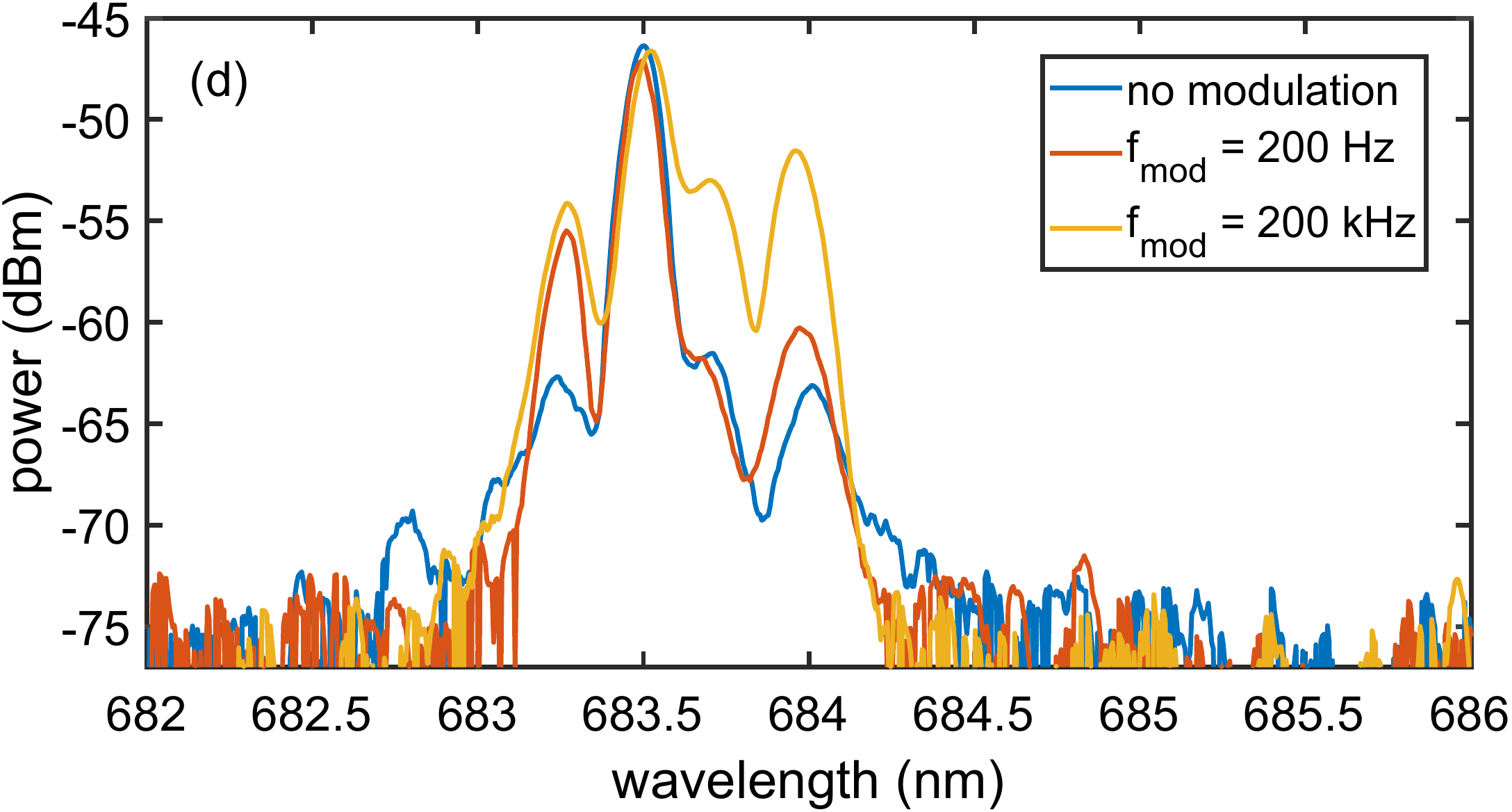}
\caption{Optical spectrum of the laser diode without (a) and with (b) optical feedback, as a function of the pump current. (c) Spectra without and with optical feedback taken at a pump current of 60~mA. (d) Spectra without and with current modulation (using a different laser diode). \label{fig3}}
\end{figure}

In Fig.~\ref{fig3}, panels (a) and (b) show in color code the optical spectra of the laser diode as a function of the pump current without and with optical feedback, respectively. Without feedback we see one dominant mode at all pump currents, whose wavelength red shift as the pump current increases. At certain pump currents we observe
mode hopping, i.e. jumps between longitudinal lasing modes. The wavelength shift and mode hopping are known to be due to thermal effects (Joule heating as the pump current increases). With optical feedback, there is a spectral broadening. As shown in Fig.~\ref{fig3}(c) the spectral width (full width half maximum, FWHM) increases from 0.1~nm (no feedback) to 1.5~nm (with optical feedback). Several lasing modes have similar power and the red shift with increasing  pump current is not as visible. The broadening of the spectrum induced by pump current modulation, shown in Fig.~\ref{fig3}(d), is very small and only slightly affects the height of the side modes; it is not measurable in terms of a significant variation of the FWHM. 

\begin{figure}[!t]
\includegraphics[width=1.0\columnwidth]{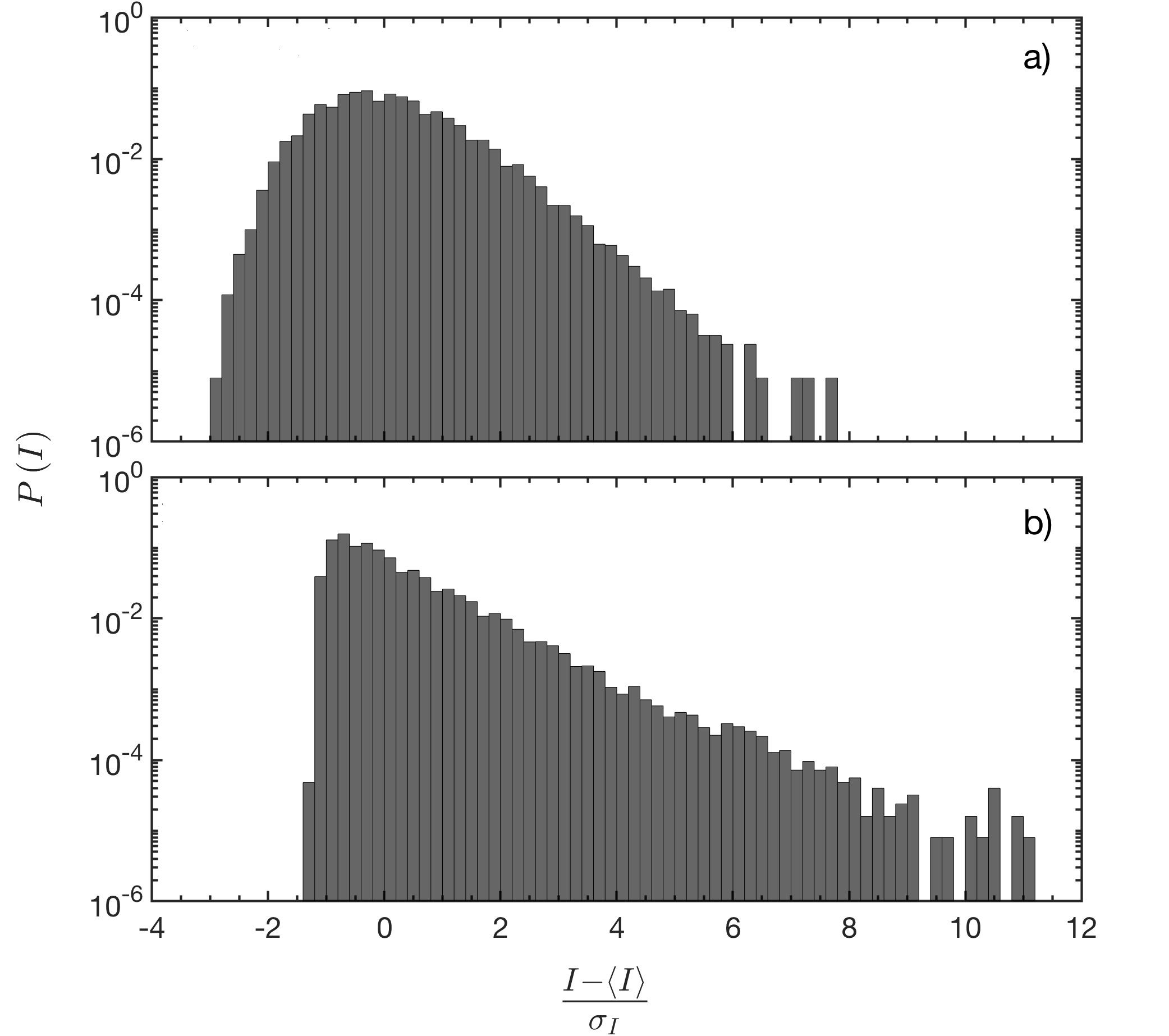}
\caption{Distribution of pixel intensities for the feedback experiment: (a) laser with optical feedback; (b) free-running laser (without optical feedback). For easier comparison the horizontal axis is normalized to zero-mean and unit variance. \label{fignewfeed}}
\end{figure}

\begin{figure}[!t]
\includegraphics[width=1.0\columnwidth]{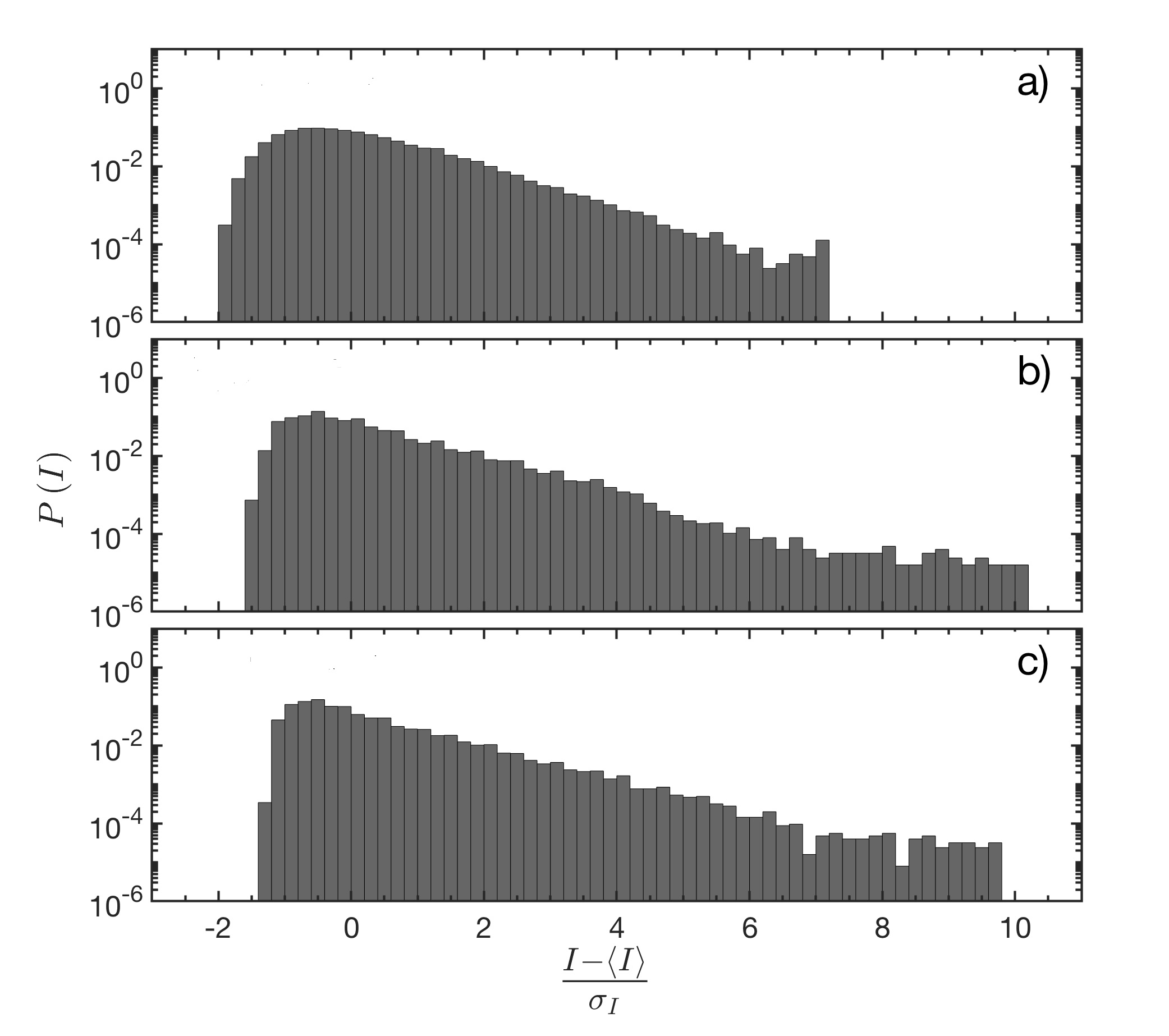}
\caption{Distribution of pixel intensities for the current modulation experiment. The  modulation frequency is (a) 200 kHz; (b) 200 Hz. Panel (c) shows the distribution for the free-running laser (i.e., without current modulation). For easier comparison the horizontal axis is normalized to zero-mean and unit variance. \label{fignewmod}}
\end{figure}

We measured the speckle contrast in a circular area in the center of the speckle image (with a radius of 200 pixels). In this way we ensure that the variability of the mean intensity across the analyzed area is low, and intensity variations are mainly due to the speckles. The distribution of intensity pixel values in this region is shown in Figs.~\ref{fignewfeed} and~\ref{fignewmod} . We see that either with optical feedback (Fig.~\ref{fignewfeed}) or with current modulation (Fig.~\ref{fignewmod}) the distribution, as expected, decays exponentially, and we note that feedback or modulation mainly modify the likelihood of small pixel values. Rayleigh statistics (correspondingly, a negative exponential intensity distribution of the speckle pattern) is to be expected under ``normal'' scattering conditions (when the total field is the sum of a large number of waves with independent amplitudes and phases, which are uniformly distributed in $0-2\pi$); however, non-Rayleigh statistics (including heavy tails) have been observed under particular scattering conditions (see, for example, Ref.\cite{cao_prl} and references therein).

When calculating the speckle contrast, we find that the speckle contrast is reduced from $C = 0.44$ (no feedback) to $C = 0.18$ (with optical feedback). Therefore, a significant reduction of the speckle contrast is produced by the light with a feedback-broadened spectrum. These results are in good agreement with early work that demonstrated feedback-induced speckle reduction~\cite{ol_1993}.

In spite of the fact that the broadening of the spectrum induced by current modulation is very small [see Fig.~\ref{fig3}(d)], when measuring the speckle contrast in a circular area as in the feedback experiment, we observe a reduction from $C = 0.53$ for no-modulation, to $C = 0.41$ for a modulation frequency of 200~Hz, and to $C = 0.34$ for a modulation frequency of 200~kHz.

Since the observed spectral broadening is small, but the speckle reduction is rather larger (with optical feedback: 58\%; with pump current modulation: 23\% or 36\% for a modulation frequency of 200 Hz or 200 kHz), we speculate that there are underlying fast and large variations of the optical spectrum, which are responsible for the speckle reduction, but which can not be detected by our detection system, mainly because the optical spectrum analyzer is very slow in comparison with the time scale of the spectral variations (for high spectral resolution the recording time of the OSA is of the order of 10s, while the characteristic mode-switching time is of few $\mu$s\cite{tredicce}).

\section{Nanolasers setup}

The fabrication and characteristics of the nanolasers where described in~\cite{nano1,nano2}. They consist of two evanescently coupled active photonic crystal L3 cavities (three holes missing) in a semiconductor free-standing membrane. They have two coexisting modes (the B=bonding mode is a symmetric superposition of single cavity modes while the AB=anti-bonding mode is an anti-symmetric superposition). 

The experimental setup is shown in~\cite{nano2}, supplementary information. The nanolasers are mounted on a piezo stage so that they can be precisely positioned relatively to the pump beam --a 100 ps-pulse semiconductor laser (Delta Diode DD-785L)-- which is focused by a X100 microscope objective. This objective is also used to collect the light and guided it to two identical low noise avalanche photodetectors (APDs, Princeton Lightwave PLA-841-FIB) as well as to a spectrometer or to an infrared camera (Sensors Unlimited SU 320). Two single-mode fibers coupled to microscope objectives are used to spatially select two regions of the far field: the center corresponds to B mode intensity, and one of the lateral lobes to the AB mode. These optical signals are sent to the APDs for monitoring simultaneous in an oscilloscope the temporal intensity fluctuations. The spectrometer allows checking the selected band before measurements (band selection by an angle-dependent spectral filter), as the bonding and anti-bonding modes are spectrally separated by just few nanometers. With the infrared camera, we can observe the near-field as well as the far-field by inserting a lens in the beam path. In order to generate a speckle pattern, a ground glass disk (GGD) and two confocal lenses in the beam path are used. The GGD scatters the light and imposes phase differences on the beam which lead to speckle. A third lens allows imaging the speckle pattern onto the camera sensor.

\section{Nanolasers results}
\begin{figure}[!t]
\includegraphics[width=0.7\columnwidth]{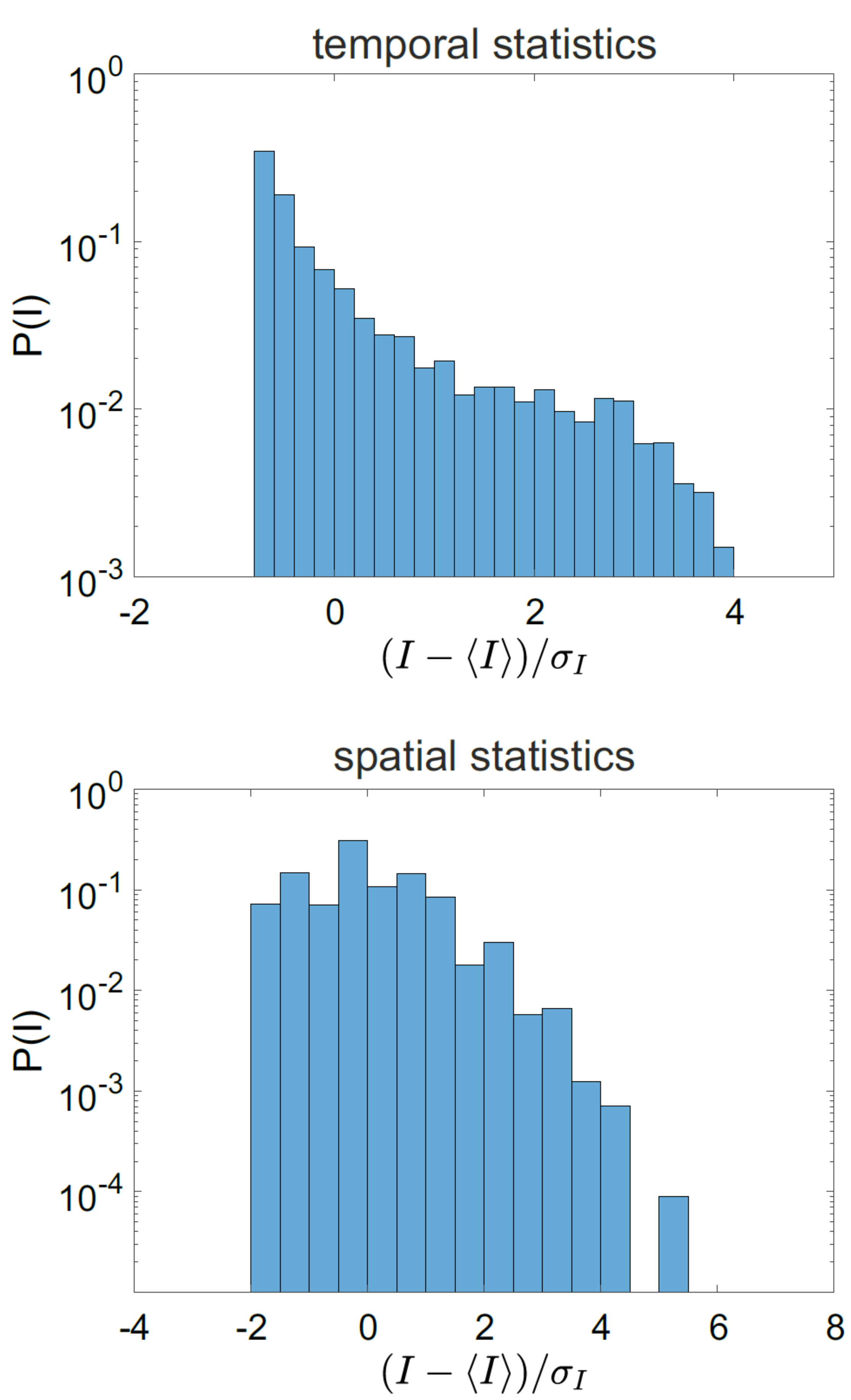}
\caption{Distribution of temporal intensity fluctuations (distribution of values in the intensity time series, top) and distribution of spatial values (pixel intensities, bottom). In both cases the horizontal axis is normalized to zero-mean and unit variance. \label{fig4}}
\end{figure}

\begin{figure}[!t]
\includegraphics[width=0.7\columnwidth]{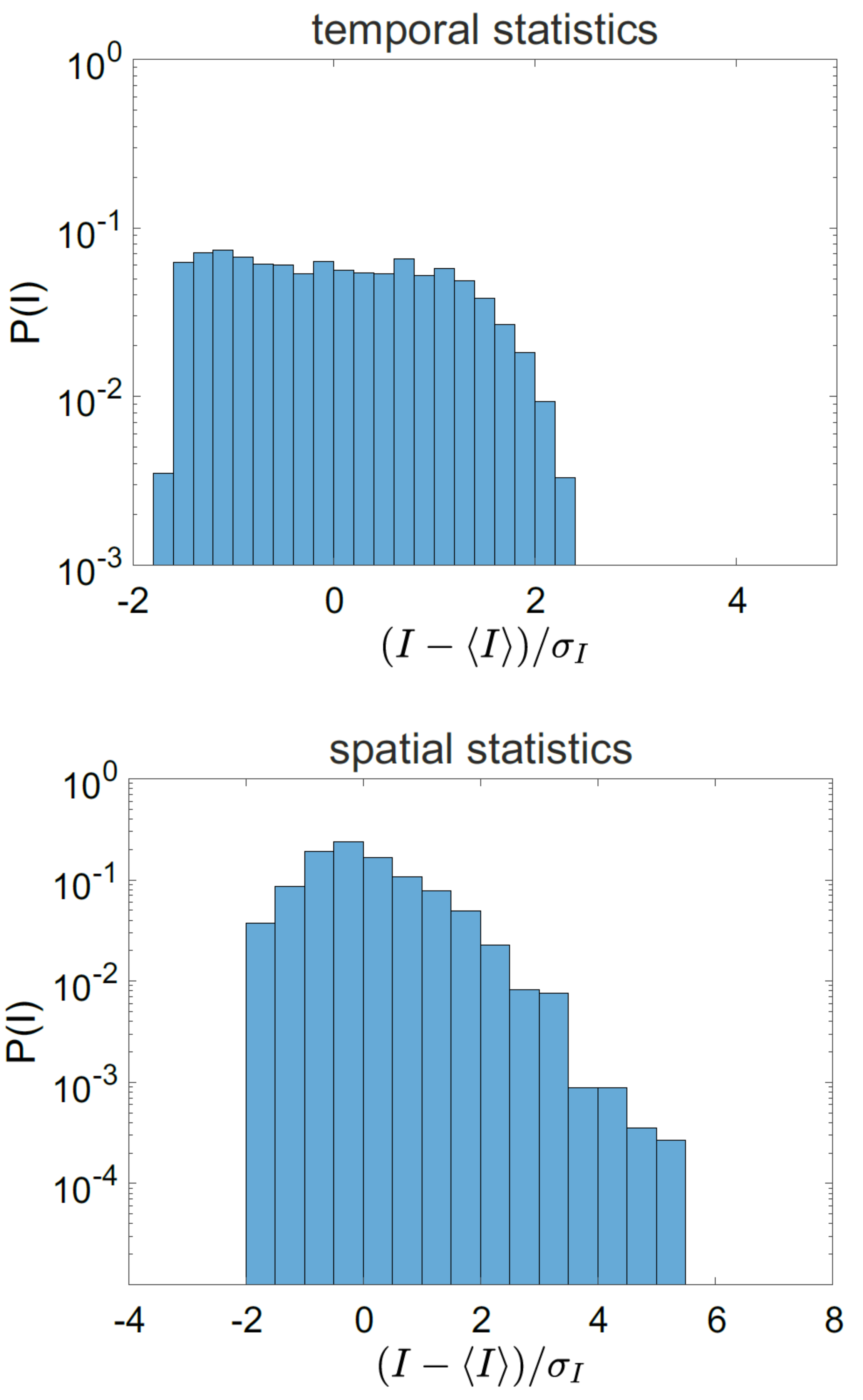}
\caption{As Fig.~\ref{fig4}, but for a different coupling of the nanolasers' modes. \label{fig5}}
\end{figure}

The results of two measurements are presented in Figs.~\ref{fig4} and \ref{fig5}, which display the probabilities of spatial and temporal intensity fluctuations of the bonding mode (B). This is the ``non-lasing'' mode of the system and thus the one that can have superthermal intensity fluctuations. The differences in the temporal and spatial intensity distributions are due to a slight change of the position of the pump spot in y-direction (realized using the piezo stage), which in turn changes the coupling of the modes of the coupled nanolasers. In Fig.~\ref{fig4} the temporal fluctuations have super-thermal statistics as the distribution of values in the intensity time series is long tailed. In contrast, in Fig.~\ref{fig5} the temporal fluctuations have thermal statistics as the distribution has a sharp cut off. The generated speckle pattern is, in both cases, consistent with Rayleigh statistics; however, we observe a change in the shape of the spatial histogram: when the temporal statistics is thermal, the distribution of intensity pixels has a rather clear exponential shape, while, when the temporal statistics is non-thermal, the shape is more irregular. In spite of the different shape, the speckle constrast is similar in the two cases (C=0.59 in Fig.~\ref{fig4} and C=0.56 in Fig.\ref{fig5}).  We speculate that the camera's relatively low resolution (or, equivalently, the very low optical power emitted by the nano lasers) is the main problem for detecting significant differences in the value of the speckle contrast. 


\section{Discussion and conclusions}

As explained in the Introduction, speckle patterns can be used to infer information about the scattering medium (if the spectral properties of the light are known), or to extract information of the properties of the light source (if the scattering medium is known). Here we have taken this second approach and we have used speckle (generated due to propagation along a multimode fiber, or through a ground glass disk) as a tool for gaining insight into the coherence of semiconductor laser light. 

In the first part of this work, using a conventional diode laser, we have observed a significantly large reduction of speckle produced by optical feedback or by pump current modulation. With both techniques, we observed only a small broadening of the optical spectrum. The speckle reduction was interpreted as due to fast and large variations of the optical spectrum, which can not be resolved by our detection system. Future work will be aimed at mapping the speckle reduction as a function of the current modulation parameters (pump current dc value, modulation amplitude and frequency), of the optical feedback parameters (feedback strength, delay time), and also, of the camera exposure time.

In the second part of this work we have analyzed speckle patterns generated by the light emitted by two coupled nanolasers, using a setup that allows modifying the coupling between two eigenmodes of the coupled system, which in turn results in the emitted light having different temporal statistics. We found that both, thermal and non-thermal light generated speckle patterns which, while showing some differences between them, were both consistent with normal Rayleigh statistics. The lack of non-Rayleigh statistics could either be due to the fact that the speckle images have very low intensity values (high signal to noise ratio), or due to the fact that temporal fluctuations are too fast and render the speckle pattern almost effectively uncoupled from the temporal intensity dynamics. Future work is aimed at clarifying this point.

Taken together, our results show that speckle patterns provide a non-spectral way to assess the coherence of semiconductor laser light.

\begin{acknowledgments}
This work was supported in part by Spanish Ministerio de Ciencia, Innovación y Universidades  (PGC2018-099443-B-I00), ITN BE-OPTICAL (H2020-675512) and ICREA ACADEMIA (C. M.), Generalitat de Catalunya.
\end{acknowledgments}

\section*{DATA AVAILABILITY}
The data that supports the findings of this study are available within the article.

\end{document}